\documentclass[useAMS,usenatbib,usegraphicx]{mn2e}
\usepackage{amsmath,times}

\newtheorem{definition}{Definition}
\newcommand{\ud}{\mathrm{d}}
\newcommand{\ue}{\mathrm{e}}
\newcommand{\ui}{\mathrm{i}}

\title%
{Foreground contamination of the WMAP CMB maps
from the perspective of the matched circle test}
\author[H.~Then]{H.~Then$^{1}$\thanks{E-mail: holger.then@uni-ulm.de}\\
$^{1}$Abteilung Theoretische Physik, Universit\"{a}t Ulm,
 Albert-Einstein-Allee~11, 89069~Ulm, Germany}
\begin{document}

\date{
  Received 2006 July 12; in original form 2005 November 25 }
\pagerange{\pageref{firstpage}--\pageref{lastpage}} \pubyear{2006}
\maketitle
\label{firstpage}

\begin{abstract}
  {\it WMAP\/} has provided CMB maps of the full sky.  The raw data is
  subject to foreground contamination, in particular near to the
  Galactic plane.  Foreground cleaned maps have been derived, e.g.,
  the internal linear combination (ILC) map of \citet{Bennett2003},
  and the reduced foreground TOH map of \citet{Tegmark2003}.  Using
  $S$ statistics we examine whether residual foreground contamination
  is left over in the foreground cleaned maps.  In particular, we
  specify which parts of the foreground cleaned maps are sufficiently
  accurate for the circle-in-the-sky signature.  We generalise the $S$
  statistic, called $D$ statistic, such that the circle test can deal
  with CMB maps in which the contaminated regions of the sky are
  excluded with masks.
\end{abstract}

\begin{keywords}
  methods: data analysis -- methods: statistical -- cosmic microwave
  background -- large-scale structure of Universe -- cosmology:
  miscellaneous
\end{keywords}

\section{INTRODUCTION} \label{sec:1} Astronomical observations of
recent years have answered a number of fundamental questions.  A
cornerstone in {\em revealing the state of the universe} was the
combination of observations at low redshift (clusters, including the
mass-to-light method, baryon fraction, and cluster abundance
evolution), intermediate redshift (SNe), and high redshift (CMB)
\citep{Bahcall1999}.  On the other hand, many new fundamental
questions were raised, e.g., questions concerning dark matter and dark
energy.  Moreover, many answers depend sensitively on the accuracy of
the observations, e.g., on the accuracy of the curvature which
determines whether we live in an {\em exactly} flat universe or a
slightly curved one.

Concerning the high redshift observations, the temperature
fluctuations in the CMB are subject to systematic errors resulting
from foreground contamination.  In particular near to the Galactic
plane, the CMB maps are highly contaminated by radiation from the
Milky Way.  Fortunately, there exist methods to reduce the foreground
contamination in the maps \citep{Tegmark1996,Bennett2003}.  A growing
number of foreground cleaned maps are available
\citep*{Bennett2003,Tegmark2003,Eriksen2004,Hinshaw2006} and the
reader may ask which of them comes closest to the genuine temperature
fluctuations of the CMB.  This question is related to the question how
reliable the maps are, especially near to the Galactic plane.  There
is no way to decide whether a given pixel of a map displays the
correct temperature (apart from pixels that are really very off in
their values), because the genuine temperature fluctuations of the CMB
are unknown.  But taking ensembles of pixels, it is possible to check
statistically whether the spatial temperature distribution agrees with
statistical isotropy.  In this paper we perform such checks using $S$
statistic and a generalisation of it.  The result is that even for the
best available CMB map either there is left residual foreground
contamination or the angular resolution is significantly reduced.
Because of this residual foreground contamination, we extend the
definition of the $S$ statistic such that the foreground contaminated
regions of the sky can be excluded.  For further studies, this might
open a new chance for the detection of matched circles as a result of
the non-trivial topology of the universe.

\section{\textbfit{D$\,$} STATISTIC} \label{sec:2} $S$ statistic was
initially introduced by \citet*{Cornish1998} for the search of
correlated pairs of circles in the CMB sky,
\begin{align}
  S=\frac{\langle 2\delta T_1(\pm\phi)\delta T_2(\phi+\phi_{*})
    \rangle}{\langle \delta T_1(\phi)^2 + \delta T_2(\phi)^2 \rangle},
  \label{eq:2.1}
\end{align}
where $\delta T_i(\phi)$ $\  (i=1,2)$ denotes the CMB temperature
fluctuations on two circles 1 and 2, and $\langle\  \rangle =
\int_{0}^{2\upi}\ud\phi$ is the integration along the two circles of
equal radius with relative phase $\phi_{*}$.

We generalise the definition of the $S$ statistic to include pixel
weights that specify the certainty resp.\  uncertainty of the
temperature fluctuations.  The new statistic is called $D$ statistic.
In addition, we note that the $D$ statistic is not restricted to
correlations along circles, but can be used along arbitrary curves
$\gamma$.

\begin{definition}
  \label{def:1} Let $\gamma_1$ and $\gamma_2$ be curves parametrized
  by $\phi$, $\  \phi\in[0,2\upi]$, and $w(\gamma_i)$ be positive pixel
  weights that specify how accurate the CMB temperature fluctuations
  $\delta T(\gamma_i)$ are known at $\gamma_i=\gamma_i(\phi)$.  Using
  the notation
  \begin{multline*}
    \langle\langle f(\gamma_1,\gamma_2) \rangle\rangle :=
    \int_{0}^{2\upi}\ud\phi\,f\big(\gamma_1(\phi),\gamma_2(\phi)\big)
    \times \\  \times \sqrt{ w\big(\gamma_1(\phi)\big)
      w\big(\gamma_2(\phi)\big)
      \big|\tfrac{\ud}{\ud\phi}\gamma_1(\phi)\big|
      \big|\tfrac{\ud}{\ud\phi}\gamma_2(\phi)\big| },
  \end{multline*}
  we call
  \begin{align}
    D := \frac{\langle\langle 2\delta T(\gamma_1)\delta T(\gamma_2)
      \rangle\rangle}{\langle\langle \delta T(\gamma_1)^2 + \delta
      T(\gamma_2)^2 \rangle\rangle} \label{eq:2.3}
  \end{align}
  $D$ statistic.
\end{definition}

We remark that the $D$ statistic is invariant if the weights are
rescaled, i.e., $w(\gamma_i(\phi)) \mapsto c w(\gamma_i(\phi))$, where
$c$ is a positive constant.  The same is also true if the temperatures
are rescaled.

The pixel weights may be expressed by $w=(1+1/\xi)^{-1}$, where $\xi$
is the signal-to-noise ratio.  If a temperature fluctuation is known
exactly, the corresponding weight equals 1.  In the other extreme,
i.e., when $\delta T$ is not known at all, its weight vanishes.

It is allowed to parametrise the curves $\gamma_i$ arbitrarily, but
care has to be taken if the ratio between
$|\frac{\ud}{\ud\phi}\gamma_1(\phi)|$ and
$|\frac{\ud}{\ud\phi}\gamma_2(\phi)|$ varies with $\phi$, because the
$D$ statistic depends on the parametrization.  We will always choose
the parametrization such that $|\frac{\ud}{\ud\phi}\gamma_i(\phi)| =
\text{\it const}$ allowing us to drop the factors
$|\frac{\ud}{\ud\phi} \gamma_i(\phi)|$ in the definition of the $D$
statistic.

If one chooses the curves $\gamma_1$ and $\gamma_2$ to be circles of
the same radius and sets all pixel weights equal to 1, one recovers
(\ref{eq:2.1}) as a special case of (\ref{eq:2.3}).

The $D$ statistic always takes values between $D=-1$, maximal {\em
  anticorrelation}, and $D=1$, maximal {\em correlation}.  In a
typical resolution-limited application the $D$ statistic has a value
distribution that is centred near
\begin{align}
  D_{\text{peak}} = \frac{\chi^2}{1+\chi^2} \label{eq:2.4}
\end{align}
and has
\begin{align}
  \text{FWHM} \simeq \sqrt{ \frac{8\ln2}{N_0} } \label{eq:2.5}
\end{align}
for $N_0$ large, where $N_0$ is the number of {\em independent} pixels
\citep{Cornish1998}.  $\chi$ is the correlation ratio,
\begin{align}
  \chi^2 = \frac{A_{\text{corr}}^2}{A^2-A_{\text{corr}}^2},
  \label{eq:2.6}
\end{align}
where $A_{\text{corr}}$ stands for the mean amplitude of the
correlated part of the signal, and $A$ for the mean amplitude of the
total signal including the detector noise, i.e.,
$A^2=A_{\text{corr}}^2 +A_{\text{uncorr}}^2+A_{\text{noise}}^2$.

Moreover, we also take angular weights into consideration.  In
particular, we emphasise that we always remove the DC and the lowest
frequency AC components along each curve $\gamma_i$, cf.\  Appendix
\ref{app:A}.  Especially in Sec.~\ref{sec:5}, we use the angular
weights of \citet{Cornish2004}.

\section{THE METHOD} \label{sec:3} Observing the CMB over the full
sky, one is confronted with foreground emission of our own galaxy that
strongly contaminates the temperature fluctuations near to the
Galactic plane.  A quantitative measure of this foreground
contamination can be obtained from the $D$ statistic by choosing
$\gamma_1$ to be a closed path and setting
$\gamma_2(\phi)=\gamma_1(\phi_{*}-\phi)$.  In the subcase of circles
we call this setup {\em front-to-front} circles, because $\gamma_2$
traverses the same circle as $\gamma_1$, but in the opposite
direction.

\begin{figure}
  \centering
  \includegraphics{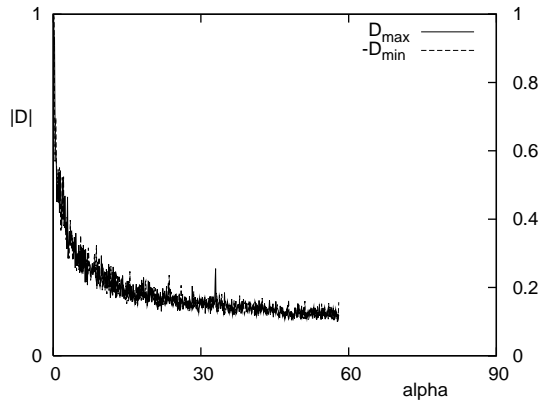} \caption{$D$ statistic of the foreground
    cleaned $W$-band map where the extremal $D$-values, i.e.\
    $D_{\text{max}}$ and $-D_{\text{min}}$, are plotted for
    front-to-front circles that are parallel to the Galactic plane.
    $\alpha$ is the circle radius.  Restricting to $\alpha<58$ degree
    the Galactic foreground contamination is not displayed.  The
    extragalactic contamination is quite neglible, since it does not
    destroy the symmetry between $D_{\text{max}}$ and
    $D_{\text{min}}$.}
  \label{fig:3.1}
\end{figure}

\begin{figure}
  \centering
  \includegraphics{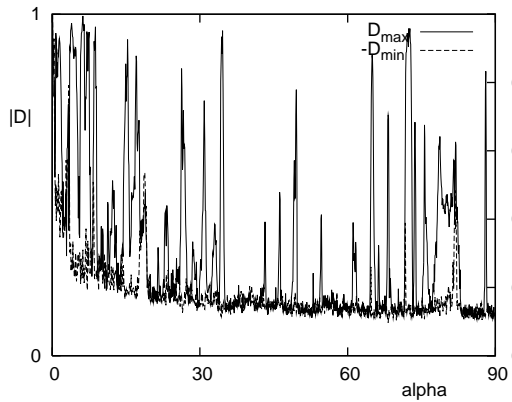} \caption{$D$ statistic of the foreground
    cleaned $W$-band map where the front-to-front circles cross the
    Galactic plane perpendicular.  The peaks in $D_{\text{max}}$ and
    the asymmetry between $D_{\text{max}}$ and $D_{\text{min}}$ result
    from Galactic foreground contamination.}
  \label{fig:3.2}
\end{figure}

Our working hypothesis is that for almost all points on the sky there
are temperature fluctuations in the CMB on all scales and that the
cosmological principle holds.  Consequently, we do not expect any
correlations in the genuine temperature fluctuations, i.e.\
$A_{\text{corr}}=0$, apart from the circle-in-the-sky signature in
case of a non-trivial topology of the universe.  In other words, the
typical temperature fluctuations of the CMB should result in a $D$
statistic that is symmetric around $D_{\text{peak}}=0$.  That this
expectation indeed holds is shown with front-to-front circles that are
far away from the Galactic plane, see Fig.~\ref{fig:3.1}.
(All figures in this section, namely
Figs.~\ref{fig:3.1}--\ref{fig:3.3}, were made with the foreground
cleaned $W$-band map of the first year {\it WMAP\/} data
\citep{Bennett2003}.  More details about the CMB maps can be found in
Sec.~\ref{sec:4}.) What we actually plot in Fig.~\ref{fig:3.1} is not
a histogram of the $D$-values, but $-D_{\text{min}}:=-\min_{\phi_{*}}
D$ and $D_{\text{max}}:=\max_{\phi_{*}} D$ in dependence of the circle
radius $\alpha$.  Being symmetric around $D_{\text{peak}}=0$, the
smallest and largest quantities of a finite set of $D$-values have
similar size, except of their opposite signs, i.e.,
$-D_{\text{min}}\approx D_{\text{max}}$.  In short, we call this to be
a {\em symmetry} between $D_{\text{min}}$ and $D_{\text{max}}$.  In
Fig.~\ref{fig:3.1} we see also that the width of the distribution
which is related to the sum of $-D_{\text{min}}+D_{\text{max}}$
decreases monotonically with increasing radius $\alpha$ of the
circles, apart from statistical fluctuations.  This is in agreement
with Eq.~(\ref{eq:2.5}), since larger circles sample more independent
pixels.  On the other hand, front-to-front circles that cross the
Galactic plane perpendicular reveal strong correlations resulting from
foreground contamination, see Fig.~\ref{fig:3.2}.  Both, the symmetry
between $D_{\text{min}}$ and $D_{\text{max}}$, and the monotonically
decreasing sum $-D_{\text{min}}+D_{\text{max}}$ are heavily distorted
by peaks in $D_{\text{max}}$ that result from residual foregrounds.
Figure \ref{fig:3.3} looks completely different to the former.  The
only difference in producing it, was to apply a galaxy cut.  While in
Fig.~\ref{fig:3.2} all pixel weights were set equal to 1, in
Fig.~\ref{fig:3.3} those inside the Kp4 mask of \citet{Bennett2003}
were set to zero which is tantamount to taking only those parts of the
circles that are outside the galaxy cut.  The symmetry between
$D_{\text{min}}$ and $D_{\text{max}}$, and the monotonically
decreasing width of the distribution in Fig.~\ref{fig:3.3} show that
there is far less contamination outside the Kp4 mask.  The
Figs.~\ref{fig:3.2}--\ref{fig:4.5} were all made with the same
front-to-front circles, i.e., circles that cross the Galactic plane
perpendicular.  It is therefore possible to compare these figures
directly with each other.

Of interest to the current paper are also the other foreground reduced
maps, i.e., the ILC, the LILC, and the two TOH maps.  We question how
reliable these maps are, in particular near to the Galactic plane.
With reliable we mean whether there are any correlations from
foreground contamination or any systematics from the process of
creating the maps.  We emphasize that {\em not all} foreground
systematics necessarily result in correlations.  For example
uncorrelated Gaussian noise that contaminates the data will never be
recovered with the $D$ statistic.  But turning the tables, as soon as
we find any correlation, we doubt that the temperature fluctuations
are free of residual foreground systematics.  Applying masks allows us
to locate the foreground contaminated regions.  The only exception are
correlations according to the circle-in-the-sky signature.  If
present, one can circumvent the latter in choosing paths that are not
circles.

\section{CMB MAPS} \label{sec:4} The following foreground reduced
first year CMB maps are available from the LAMBDA archive:
\begin{enumerate}
\item The foreground cleaned $Q$-, $V$-, and $W$-bands of the first
  year {\it WMAP\/} data \citep{Bennett2003} that result from removing
  the free-free, synchrotron, and dust emissions via externally
  derived CMB template fits.  The advantages of these maps are the
  high resolution, in particular 0.23 degree for the $W$-band, the
  well-specified noise properties, and the frequency specific
  information that is contained in the three microwave bands, Q, V,
  and W.
\item The internal linear combination (ILC) map that is designed to
  reduce the foreground contamination via weighted combinations of the
  five {\it WMAP\/} bands \citep{Bennett2003}.
\item The LILC map of \citet{Eriksen2004} produced with a variant of
  the ILC algorithm that employs a Lagrange multiplier.
\item The reduced foreground TOH map of \citet{Tegmark2003} that
  results from a variant of the \citet{Tegmark1996} technique which
  makes no assumptions about the CMB power spectrum, the foregrounds,
  the {\it WMAP\/} detector noise or external templates.  Since the
  TOH map is free of assumptions about external templates, it has the
  advantage that it can be used for cross-correlation with, e.g.\
  galaxy and $X$-ray maps.  Furthermore, according to
  \citet{Tegmark1996} their technique preserves the information on the
  noise properties of the CMB map.  In principle, \citet{Tegmark2003}
  should be able to specify the noise properties of their map.  But
  unfortunately, there are no noise properties given for the TOH map.
\item The Wiener filtered TOH map \citep{Tegmark2003} which is
  designed for visualisation purposes in order to represent the best
  guess as to what the CMB sky actually should look like.
\end{enumerate}
Applying the $D$ statistic to these maps, we find:
\begin{figure}
  \centering
  \includegraphics{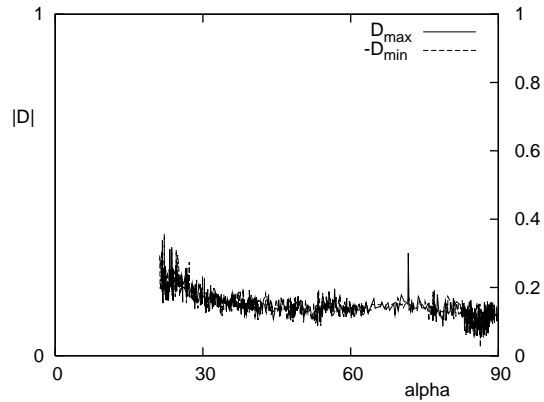} \caption{The same as in
    Fig.~\ref{fig:3.2}, except that the Kp4 mask has been applied.
    This shows that there is far less foreground contamination outside
    the Kp4 mask.  Circles with $\alpha<21$ degree are not displayed,
    because a too large fraction of each of them is inside the Kp4
    mask.}
  \label{fig:3.3}
\end{figure}
\begin{enumerate}
\item The foreground cleaned $Q$-, $V$-, and $W$-band maps are useful
  for quantitative analyses outside the Kp4 mask, see
  Fig.~\ref{fig:3.3} and the text in Sec.~\ref{sec:3}.  Inside the Kp2
  mask these maps should only be used with care, because the templates
  were fit with data outside the Kp2 mask and the foreground removal
  is not applicable to regions near to the Galactic plane where the
  spectrum of the Galactic emission is different.  In particular, the
  maps should never be used inside the Kp4 mask, see
  Fig.~\ref{fig:3.2}.
\item According to \citet{Bennett2003} the ILC map is useful for
  visual presentation of the CMB anisotropy signature and for
  foreground studies.  However, because of the complicated noise
  correlations, it should not be used for CMB studies.  Our finding is
  that the ILC map is not this bad.  Only the distribution of the $D$
  statistic is very wide, see Fig.~\ref{fig:4.1}.  The large values of
  $-D_{\text{min}}$ and $D_{\text{max}}$ highlight the low resolution
  of the ILC map, cf.\  Eq.~(\ref{eq:2.5}) and the text in
  Sec.~\ref{sec:3}.  Nevertheless, the ILC map has the advantage that
  it does not require any galaxy cut to be applied.  In order to find
  a symmetry between $D_{\text{min}}$ and $D_{\text{max}}$ for the ILC
  map, it is indispensable that the DC component along the
  front-to-front circles was removed in our analyses.  If one would
  compute the $D$ statistic of the ILC map with the DC component
  included, the symmetry between $D_{\text{min}}$ and $D_{\text{max}}$
  would be distorted.  This indicates some bias in the ILC map with
  respect to the DC component along front-to-front circles.
\item The LILC map shows similar results as the ILC map, see
  Fig.~\ref{fig:4.2}.
\item The $D$ statistic reveals that the TOH contains some residual
  foreground systematics if one analyses the full sky without any cut,
  see Fig.~\ref{fig:4.3}.  These foreground systematics require the
  Kp4 mask to be applied, see Fig.~\ref{fig:4.4}.  If compared to the
  foreground cleaned $W$-band map of \citet{Bennett2003}, see
  Fig.~\ref{fig:3.3}, the distribution of the $D$ statistic is
  slightly wider for the TOH map indicating that the resolution is
  slightly worse than the claimed 13 arcmin.  There is one sharp peak
  in Fig.~\ref{fig:4.4} at $\alpha=71.7$ degree.  Less pronounced,
  this peak occurs also in Fig.~\ref{fig:3.3}.  The reader however is
  warned not to misinterpret this peak as a detection of a matched
  circle pair.  If the Kp4 mask combined with the point source mask of
  \citet{Bennett2003} is applied, the sharp peak in Fig.~\ref{fig:4.4}
  vanishes completely.  This tells us that there is some residual
  foreground contamination outside the Kp4 mask, but inside the point
  source mask.
\item The Wiener filtered TOH map shows similar results as the ILC and
  the LILC map, except that the distribution of the $D$ statistic is
  somewhat smaller which indicates that the resolution of the Wiener
  filtered TOH map is better than 1 degree.
\end{enumerate}
\begin{figure}
  \centering
  \includegraphics{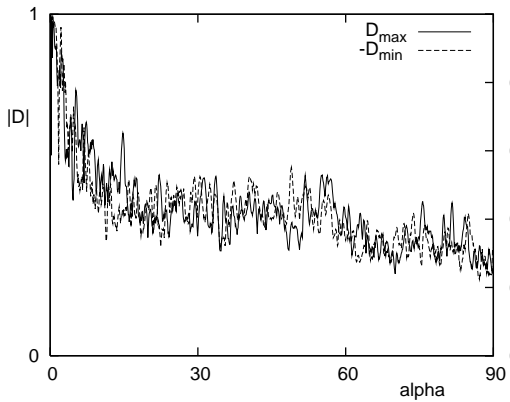} \caption{$D$ statistic of the ILC map.}
  \label{fig:4.1}
\end{figure}
\begin{figure}
  \centering
  \includegraphics{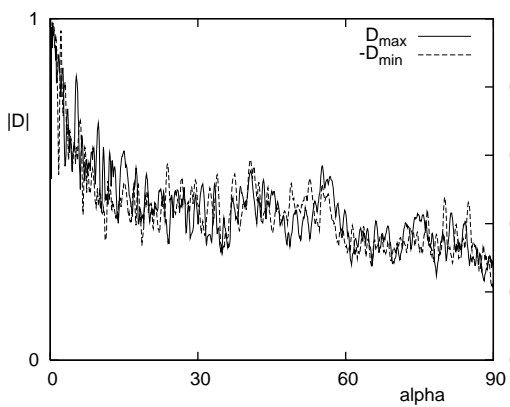} \caption{$D$ statistic of the LILC map.}
  \label{fig:4.2}
\end{figure}

\section{TOPOLOGY AND CIRCLES IN THE SKY} \label{sec:5} In the
following we choose a linear superposition of the foreground cleaned
$Q$-, $V$-, and $W$-bands of \citet{Bennett2003} as described in
\citet{Cornish2004} and use the matched circle test in order to search
for signatures of a non-trivial topology.  Such searches have already
been done for back-to-back circles
\citep*{deOliveira2004,Roukema2004,Aurich2005c} and have been extended
to nearly back-to-back circles \citep{Cornish2004}.  In order not to
swamp the SW by the ISW we use the angular weights that have been
introduced in \citep{Cornish2004}.

Since \citet{Cornish2004}, cf.\  also \citet{Shapiro2006}, have not
found any matched circle pair, one could conclude that the universe
has a trivial topology.  At this point it is worthwhile to remember
that \citet{Roukema2004} have reported a slight signal for six circle
pairs with a radius of $11\pm1$ degree in the ILC map which they
interpret as a possible hint for the left-handed dodecahedral space.
While the systematic research for matched circles in spherical spaces
carried out by \citet{Aurich2005c} has not found these circles,
\citet{Aurich2005c} however report a marginal hint for the
right-handed dodecahedron and for the right-handed binary tetrahedral
space, respectively.  In addition, there is strong evidence for a
finite universe with a small dimension, because of the anomalies of
the large scale temperature fluctuations which rule out the
concordance model with a probability of 99.996 per cent, but could be
explained by a multiply connected universe \citep{deOliveira2004}.
These anomalies are the surprisingly low quadrupole, the
quadrupole-octupole alignment, and the planar octupole.

We are not aiming to repeat a back-to-back search.  We avoid also to
run the full search that has been started by \citet{Cornish2004},
because this would require more computer resources than we could ever
afford.  Instead, we explore whether there are any matched
front-to-front circles in the sky that result from orientation
preserving isometries.  This search has never been carried out before,
and it can be done within a few days on a multiprocessor workstation.
In our search we exclude contaminated regions of the CMB sky via the
Kp4 and the point source mask by replacing the temperature
fluctuations inside these pixels with those given by the ILC map.
\begin{figure}
  \centering
  \includegraphics{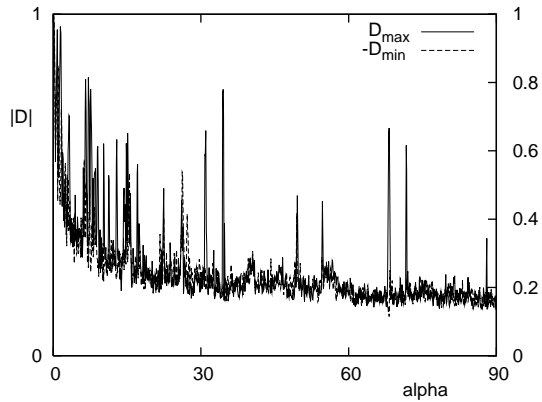} \caption{$D$ statistic of the TOH map.}
  \label{fig:4.3}
\end{figure}
\begin{figure}
  \centering
  \includegraphics{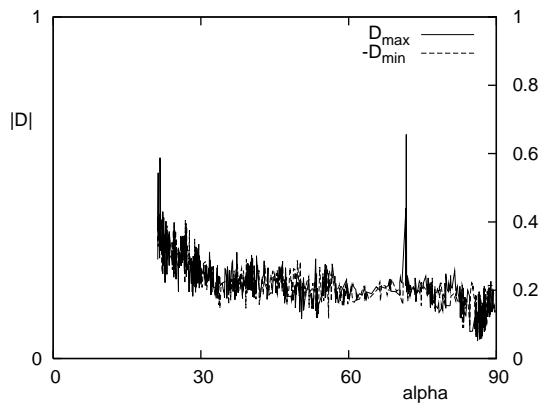} \caption{$D$ statistic of the TOH map
    outside the Kp4 mask.}
  \label{fig:4.4}
\end{figure}
\begin{figure}
  \centering
  \includegraphics{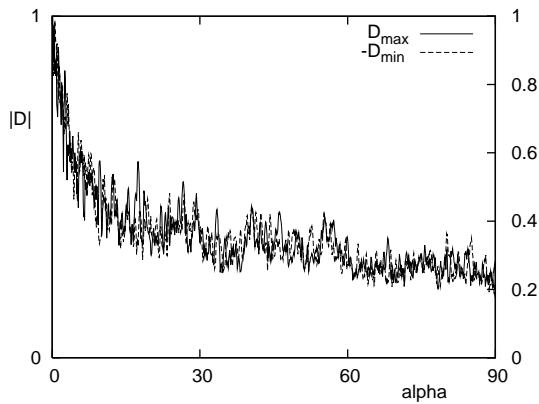} \caption{$D$ statistic of the Wiener
    filtered TOH map.}
  \label{fig:4.5}
\end{figure}

One may wonder whether it makes sense to search for front-to-front
circles.  Are there any topologies that predict front-to-front
circles? The surprising answer is {\em yes}.  An example is the Picard
universe \citep{Aurich2004}.  Choosing the cosmological parameters as
was done in \citet{Aurich2004,Aurich2005a} the Picard universe
predicts 40 circle pairs of which 23 are front-to-front.  Further
details of the Picard topology can be found in
\citet*{Aurich2003,Then2006}.  Not only the Picard, but also many
other topologies in spherical, flat, and hyperbolic space,
respectively, predict front-to-front circles, if one allows that the
multiply connected space has elliptic fixed-points.  As another
example, see the hyperbolic tetrahedral space investigated by
\cite{Aurich1999,Aurich2001}.

Exploring all front-to-front circles of orientation preserving
isometries with a HEALPIX resolution of $N_{\text{side}}=256$ we have
not found any matched circle pair that can be significantly
distinguished from false positive matches.  This leads us to the
conclusion that there are no matched front-to-front circles on the CMB
sky, but care has to be taken since this conclusion is only
significant for circles with large radius, $\alpha\gg45$ degree.
Otherwise, if $\alpha\le45$, the width of the $D$-value distribution,
cf.\  Eq.~(\ref{eq:2.5}), reaches nearly the height of the $D$-values
that are expected for matched circles, Eq.~(\ref{eq:2.4}).

In order to be more specific, we make quantitative estimates for the
width of the $D$-value distribution and for the height of the expected
$D$-values in case of matched circle pairs.  The height of the
expected $D$-values for matched circles depends sensitively on the
correlation ratio $\chi$ of Eq.~(\ref{eq:2.6}).  The correlated signal
in the CMB due to topology comes from the naive Sachs--Wolfe effect.
For a circle radius of $\alpha=45$ degree the Doppler effect
contributes to the uncorrelated part of the signal.  Any other
physical effects, e.g., the late integrated Sachs--Wolfe, and the
Sunyaev--Zeldovich effect, contribute to the uncorrelated part of the
signal, independently of the circle radius.  In addition, there are
residual foreground contamination and detector noise which are not
correlated with respect to the universal covering of the quotient
space.  Modeling the universe shows that the contribution of the late
integrated Sachs--Wolfe effect plus the contribution of the Doppler
effect is of the same order as the contribution of the naive
Sachs--Wolfe effect, see fig.~8 in \citet*{Aurich2005b}.  This results
in $A_{\text{uncorr}}^2+A_{\text{noise}}^2>A_{\text{corr}}^2$.
Consequently, the correlation ratio is $\chi<1$.  If there would be a
matched pair of circles in the sky, we would get a $D$-value for this
matched circle pair which is $D_{\text{peak}}<0.5$, see
Eq.~(\ref{eq:2.4}).  These single $D$-values, one for each matched
circle pair, are embedded in a large number of $D$-values that are
distributed around $D_{\text{peak}}=0$ with a standard deviation of
$\sigma\simeq\sqrt{1/N_0}$, cf.\  Eq.~(\ref{eq:2.5}).  The largest
number of independent pixels that a front-to-front circle can sample
in the {\it WMAP\/} data is $N_0=180\degr/0.23\degr=782$ yielding
$\sigma> 0.036$.  Due to the large number of $3\times10^{11}$
$D$-values, namely one for each circle centre times each radius times
each relative phase, many of these $D$-values reach $D_{\text{max}}\gg
5\sigma\simeq 0.18$.  Among these $D$-values, $D_{\text{max}}\gg
0.18$, it might be possible to find a matched circle that has
$D_{\text{peak}}<0.5$, but the risk that matched circles are
overlooked among many false positive detections is high.  In
particular, taking the detection threshold
\[ D_{\text{trigger}}(\alpha) \simeq \langle S^2 \rangle^{\frac{1}{2}}
\sqrt{2 \ln \Big(
  \frac{N_{\text{search}}}{2\sqrt{\pi\ln(N_{\text{search}})}} \Big) }
\] of \citet{Cornish2004} results in a number of false positive
detections for front-to-front circles, whereas there is also the risk
that even true matched circles may be missed, cf.\
\citet{Aurich2005c}.  Here, the main source of trouble is that for
front-to-front circles $D_{\text{max}}$ is larger by more than a
factor of $\sqrt{2}$ if compared to the case of back-to-back circles.

\section{THREE YEAR MAPS} \label{sec:7} In March 2006 the NASA Science
team has published updated skymaps from three years of WMAP data
collection \citep{Hinshaw2006}.  Repeating our analysis of
Sec.~\ref{sec:4} with the three year maps gives the same results as
before.  This reflects that the WMAP observations are stable over the
different years.

Also repeating the search of Sec.~\ref{sec:5} does not recover any
front-to-front circle in the three year data.

\section{CONCLUSION} \label{sec:6} Analysing CMB maps with $D$
statistics allowed us to specify whether the available maps are
contaminated by residual foregrounds.  At the same time, $D$
statistics highlighted the angular resolution of the different CMB
maps.

We conclude that, at present, the foreground cleaned $Q$-, $V$-, and
$W$-band maps of \citet{Hinshaw2006} are the best available maps of
the CMB sky.  The foregrounds in these maps have been removed with
high quality outside the Kp2 mask.  Even inside the Kp2 mask, but
outside the Kp4 and the point source mask, the foreground reduction is
very good.  We recommend to use these maps in any quantitative
analysis that allows the Kp4 mask to be applied.  If one is cautious,
it is safe to use the Kp2 mask combined with the point source mask.

Almost as well is the TOH map of \citet{Tegmark2003}.  It is reliable
outside the Kp4 and the point source mask, but its resolution is
slightly worse than the claimed 13 arcmin.  Unfortunately, the TOH map
is yet only available for the first year of WMAP data.

The ILC, the LILC, and the Wiener filtered TOH map are the only maps
that are reliable inside the Kp4 and the point source mask.
Unfortunately, because of their low resolution they do not fully meet
the requirements of the matched circle test.  Their distribution of
the $S$ and $D$ statistic is too wide.\footnote{An exception is the
  search for a given quotient space where several circle pairs are
  correlated to each other simultaneously.  For example, the
  Poincar\'e Dodecahedral space predicts six circle pairs.  In the
  simultaneous search the number of independent pixels along these six
  circle pairs is sixfold, hence narrowing the width of the
  distribution of the $S$ and $D$ statistic suitably, cf.\
  \citet{Roukema2004,Aurich2005c}.}  Nevertheless, they can be used to
replace the data of higher resolution maps inside the Kp4 and the
point source mask.

Finally, we have searched for matched front-to-front circles, but have
not found any.

\section*{ACKNOWLEDGMENTS} I thank Dr.~R.~Aurich, S.~Lustig, and
Prof.~F.~Steiner for discussions and helpful comments.  Thanks are
also to the anonymous referee for pointing out that removing the DC
component around each circle removes the spurious large correlations.
The use of the Legacy Archive for Microwave Background Data
Analysis\footnote{http://lambda.gsfc.nasa.gov/} (LAMBDA) is
acknowledged.  Support for LAMBDA is provided by the NASA Office of
Space Science.  The results in this paper have been derived using the
CFITSIO\footnote{http://heasarc.gsfc.nasa.gov/docs/software/fitsio/},
the HEALPIX\footnote{http://healpix.jpl.nasa.gov/}
\citep*{Gorski1999}, and the FFTW\footnote{http://www.fftw.org/}
\citep{Frigo2005} libraries, the \texttt{gcc} compiler and
\texttt{gnuplot}.

\appendix

\section{ANGULAR WEIGHTS} \label{app:A} Searching for the
circle-in-the-sky signature in the {\it WMAP\/} observations
\citet{Cornish2004} have included angular weights in their $S$
statistic.  Namely around each pixel $i$, they draw a circle of radius
$\alpha$ and linearly interpolate values at $n=2^{r+1}$ points along
the circle.  They then Fourier transform each circle:
$T_i(\phi)=\sum_{m}T_{im}\exp(\ui m\phi)$ and compare circle pairs:
\begin{align}
  S_{ij}(\alpha,\beta) = \frac{ 2 \hspace{-.5em}
    \sum\limits_{m=-2^r}^{2^r-1} \hspace{-.5em} |m| T_{im}(\alpha)
    T^{*}_{jm}(\alpha) \ue^{-\ui m\beta} }{
    \sum\limits_{m=-2^r}^{2^r-1} \hspace{-.5em} |m| \big[
    |T_{im}(\alpha)|^2 + |T_{jm}(\alpha)|^2 \big] }.
  \label{eq:A.2}
\end{align}
We call this to be angular weights depending on $m$.  $\beta$ is the
relative phase of the two circles and the $i,j$ label the circle
centers.  The angular weights are important, because otherwise the ISW
effect swamps the SW correlations, cf.\  \citep{Shapiro2006}.

Similarly, angular weights can be introduced for the $D$ statistic.
They enter as a multiplicative factor on the temperature fluctuations
in frequency space.

\begin{definition}
  \label{def:2} Let $f(x)$, $\hat{f}(t)$, $u(x)$, and $\hat{v}(t)$ be
  discrete functions on the points $x,\,t\in\{0,1,\ldots,n-1\}$.  Let
  \begin{align*}
    \hat{f}(t) = \mathcal{F}[f](t) = \sum_{x=0}^{n-1} f(x)
    \ue^{-2\pi\ui x t/n}
  \end{align*}
  be the Fourier transformation and
  \begin{align*}
    f(x) = \mathcal{F}^{-1}[\hat{f}](x) = \frac{1}{n} \sum_{t=0}^{n-1}
    \hat{f}(t) \ue^{2\pi\ui t x/n}
  \end{align*}
  its inverse.  We call
  \begin{align*}
    \hat{f}_{u}(t) := \frac{ \mathcal{F}[fu](t) }{ \frac{1}{n}
      \mathcal{F}[u](0) }
  \end{align*}
  the $u$ weighted Fourier transform of $f$, and
  \begin{align*}
    f_{u,\hat{v}}(x) := \frac{ \mathcal{F}^{-1}[\hat{v}\hat{f}_{u}](x)
    }{ \mathcal{F}^{-1}[\hat{v}](0) }
  \end{align*}
  the $u,\hat{v}$ weighted of the function $f$.

  The $u$ weighted Fourier transform at $t=0$ results in the
  arithmetic mean of $f$ with respect to the weights,
  \begin{align*}
    \tfrac{1}{n}\hat{f}_{u}(0) = \frac{ \sum fu }{ \sum u } =:
    \bar{f}.
  \end{align*}
  We call $\bar{f}$ to be the DC component of $f$, and
  \begin{align*}
    \tfrac{1}{n}\hat{f}_{u}(-1)\ue^{-2\pi\ui x/n}, \quad
    \tfrac{1}{n}\hat{f}_{u}(1)\ue^{2\pi\ui x/n}
  \end{align*}
  to be the lowest frequency AC components of $f$ with respect to the
  weights.
\end{definition}

We remark that the $u=1$ weighted Fourier transform of $f$ coincides
with the Fourier transform of $f$, i.e., $\hat{f}_{1}(t)=\hat{f}(t)$.
The same is true if $u(x)$ is some other non-zero constant.

The $1,1$ weighted of $f$ coincides with itself, $f_{1,1}(x)=f(x)$.
The same is true if $u(x)$ and $\hat{v}(t)$ are some other non-zero
constants.

The $u$ weights enter as a multiplicative factor in coordinate space,
\begin{align*}
  f_{u,1}(x) = \frac{ f(x) u(x) }{ \frac{1}{n} \sum u }.
\end{align*}
The $\hat{v}$ weights enter as a multiplicative factor in frequency
space,
\begin{align*}
  f_{1,\hat{v}}(x) = \frac{ \mathcal{F}^{-1}[\hat{v}\hat{f}](x) }{
    \mathcal{F}^{-1}[\hat{v}](0) } = \frac{ (v*f)(x) }{ \frac{1}{n}
    \sum \hat{v} }.
\end{align*}
If we choose $f$ to be the discretisation of the temperature
fluctuations $\delta T$ along the curve $\gamma$,
\begin{align*}
  f(x) := \delta T\big(\gamma(\phi)\big)_{| \phi=2\pi\frac{x}{n}},
  \quad x\in\{0,1,\ldots,n-1\},
\end{align*}
and $u$ to be given by the corresponding pixel weights, see
Def.~\ref{def:1},
\begin{align*}
  u(x) := \sqrt{ w\big(\gamma(\phi)\big)
    |\tfrac{\ud}{\ud\phi}\gamma(\phi)| }_{| \phi=2\pi\frac{x}{n}},
\end{align*}
we can take
\begin{align*}
  \hat{v}(t)=\hat{v}(-t), \quad t\in\{0,1,\ldots,\frac{n}{2}\} \quad
  \text{(for $n$ even)}
\end{align*}
to be real valued angular weights.

Replacing $\delta T(\gamma_i)$ $\  (i=1,2)$ by its $u,\hat{v}$ weighted
in Def.~\ref{def:1}, Eq.~(\ref{eq:2.3}), we end up with the $D$
statistic that includes both, pixel and angular weights.

If one chooses the curves $\gamma_1$ and $\gamma_2$ to be circles of
the same radius $\alpha$, sets all pixel weights equal to $w=1$, and
sets the angular weights equal to $\hat{v}(m)=\sqrt{(m)}$ for
$m=0,1,\ldots,2^r$, one recovers (\ref{eq:A.2}) as a special case of
the $D$ statistic.

When using angular weights, one has to be aware that they interact
with the angular resolution of the map in dependence of the curves
$\gamma$.  For example, smoothing the map with a Gaussian beam of root
mean square $\sigma$
is the same as multiplying the expansion coefficients $a_{lm}$ of the
map with
\begin{align*}
  \ue^{-l(l+1)\frac{\sigma^2}{2}}.
\end{align*}
A similar result happens, if one chooses $\gamma_1$ and $\gamma_2$ to
be circles of radius $\alpha$ and applies the angular weights
\begin{align*}
  \hat{v}(l)=\ue^{-l(l+1)\frac{\sigma^2}{2\sin^2\alpha}}.
\end{align*}
The difference is that the latter smoothes the map only in the
direction along the circles.

Entering in coordinate and frequency space, respectively, the pixel
and angular weights interfere.  Consequently, we avoid to use pixel
and angular weights simultaneously.  If not otherwise stated we use
pixel weights.  The only exception is that we always remove the DC and
the lowest frequency AC components of the temperature fluctuations,
i.e., we replace $\delta T$ by
\begin{align*}
  \delta T\big(\gamma(\phi)\big) - \frac{1}{n} \sum_{t=-1}^{1}
  \widehat{\delta T}_{u}(t) \ue^{\ui t\phi}
\end{align*}
in Def.~\ref{def:1}, Eq.~(\ref{eq:2.3}).

\bsp
\label{lastpage}

\begin{thebibliography}{99}
\bibitem[\protect\citeauthoryear{Aurich}{1999}]{Aurich1999} Aurich~R.,
  1999, ApJ, 524, 497
\bibitem[\protect\citeauthoryear{Aurich \& Steiner}{2001}]{Aurich2001}
  Aurich~R., Steiner~F., 2001, MNRAS, 323, 1016
\bibitem[\protect\citeauthoryear{Aurich, Steiner \& Then}{Aurich et
    al.}{2003}]{Aurich2003} Aurich~R., Steiner~F., Then~H., 2003, in
  Bolte~J., Steiner~F., eds, Proc.\  International School on
  Mathematical Aspects of Quantum Chaos II.  To appear in Lecture
  Notes in Physics, Springer--Verlag, preprint (gr-qc/0404020)
\bibitem[\protect\citeauthoryear{Aurich et al.}{2004}]{Aurich2004}
  Aurich~R., Lustig~S., Steiner~F., Then~H., 2004, Class.\  Quant.\
  Grav., 21, 4901
\bibitem[\protect\citeauthoryear{Aurich et al.}{2005a}]{Aurich2005a}
  Aurich~R., Lustig~S., Steiner~F., Then~H., 2005a, Phys.\  Rev.\
  Lett., 94, 021301
\bibitem[\protect\citeauthoryear{Aurich, Lustig \& Steiner}{Aurich et
    al.}{2005b}]{Aurich2005b} Aurich~R., Lustig~S., Steiner~F., 2005b,
  Class.\  Quant.\  Grav., 22, 2061
\bibitem[\protect\citeauthoryear{Aurich, Lustig \& Steiner}{Aurich et
    al.}{2005c}]{Aurich2005c} Aurich~R., Lustig~S., Steiner~F., 2005c,
  MNRAS, 369, 240
\bibitem[\protect\citeauthoryear{Bahcall et al.}{1999}]{Bahcall1999}
  Bahcall~N.~A., Ostriker~J.~P., Perlmutter~S., Steinhardt~P.~J.,
  1999, Science, 284, 1481
\bibitem[\protect\citeauthoryear{Bennett et al.}{2003}]{Bennett2003}
  Bennett~C.~L.\  et al., 2003, ApJS, 148, 97
\bibitem[\protect\citeauthoryear{Cornish, Spergel \& Starkman}{Cornish
    et al.}{1998}]{Cornish1998} Cornish~N.~J., Spergel~D.~N.,
  Starkman~G.~D., 1998, Class.\  Quant.\  Grav., 15, 2657
\bibitem[\protect\citeauthoryear{Cornish et al.}{2004}]{Cornish2004}
  Cornish~N.~J., Spergel~D.~N., Starkman~G.~D., Komatsu~E., 2004,
  Phys.\  Rev.\  Lett., 92, 201302
\bibitem[\protect\citeauthoryear{de Oliveira-Costa et
    al.}{2004}]{deOliveira2004} de~Oliveira-Costa~A., Tegmark~M.,
  Zaldarriaga~M., Hamilton~A., 2004, Phys.\  Rev.\  D, 69, 063516
\bibitem[\protect\citeauthoryear{Eriksen et al.}{2004}]{Eriksen2004}
  Eriksen~H.~K., Banday~A.~J., G\'{o}rski~K.~M., Lilje~P.~B., 2004,
  ApJ, 612, 633
\bibitem[\protect\citeauthoryear{Frigo \& Johnson}{2005}]{Frigo2005}
  Frigo~M., Johnson~S.~G., 2005, Proc.\  IEEE 93, 216
\bibitem[\protect\citeauthoryear{G\'{o}rski, Hivon \&
    Wandelt}{G\'{o}rski et al.}{1999}]{Gorski1999} G\'{o}rski~K.~M.,
  Hivon~E., Wandelt~B.~D., 1999, in Banday~A.~J., Sheth~R.~K.,
  Da~Costa~L., eds, Proc.\  MPA/ESO Conf., Evolution of Large-Scale
  Structure.  PrintPartners Ipskamp, NL, p.\  37
\bibitem[\protect\citeauthoryear{Hinshaw et al.}{2006}]{Hinshaw2006}
  Hinshaw~G.\  et al, 2006, preprint (astro-ph/0603451)
\bibitem[\protect\citeauthoryear{Roukema et al.}{2004}]{Roukema2004}
  Roukema~B.~F., Lew~B., Cechowska~M., Marecki~A., Bajtlik~S., 2004,
  A\&A, 423, 821
\bibitem[\protect\citeauthoryear{Shapiro Key et
    al}{2006}]{Shapiro2006} Shapiro~Key~J., Cornish~N.~J.,
  Spergel~D.~N., Starkman~G.~D., 2006, preprint (astro-ph/0604616)
\bibitem[\protect\citeauthoryear{Tegmark \&
    Efstathiou}{1996}]{Tegmark1996} Tegmark~M., Efstathiou~G., 1996,
  MNRAS, 281, 1297
\bibitem[\protect\citeauthoryear{Tegmark, de Oliveira-Costa \&
    Hamilton}{Tegmark et al.}{2003}]{Tegmark2003} Tegmark~M.,
  de~Oliveira-Costa~A., Hamilton~A.~J.~S., 2003, Phys.\  Rev.\  D, 68,
  123523
\bibitem[\protect\citeauthoryear{Then}{2006}]{Then2006} Then~H., 2006,
  in Cartier~P., Julia~B., Moussa~P., Vanhove~P., eds, Frontiers in
  Number Theory, Physics, and Geometry.  Springer--Verlag, Berlin, p.\
  183
\end{thebibliography}
\end{document}